# Focusing Surface Acoustic Waves with a Plasmonic Hypersonic Lens

Hilario D. Boggiano[1], Lin Nan[2], Gustavo Grinblat[1,3], Stefan A. Maier[4,5], Emiliano Cortés[2], and Andrea V. Bragas*[1,3]

[1] Universidad de Buenos Aires, Facultad de Ciencias Exactas y Naturales, Departamento de Física. 1428 Buenos Aires, Argentina.

[2] Chair in Hybrid Nanosystems, Nanoinstitute Munich, Faculty of Physics, Ludwig-Maximilians-Universität München, 80539 München, Germany

[3] CONICET - Universidad de Buenos Aires, Instituto de Física de Buenos Aires (IFIBA). 1428 Buenos Aires, Argentina.

[4] School of Physics and Astronomy, Monash University, Clayton, VIC 3800, Australia

[5] Department of Physics, Imperial College London, London SW7 2AZ, UK

* bragas@df.uba.ar

**Abstract**

Plasmonic nanoantennas have proven to be efficient transducers of electromagnetic to mechanical energy and vice versa. The sudden thermal expansion of these structures after an ultrafast optical pulsed excitation leads to the emission of hypersonic acoustic waves to the supporting substrate, which can be detected by another antenna that acts as a high-sensitive mechanical probe due to the strong modulation of its optical response. Sophisticated fabrication techniques, together with the implementation of numerical simulations, have allowed the engineering of nanostructures for the controlled directional generation and detection of high-frequency acoustic phonons at the nanoscale, with many potential applications in telecommunications, sensing, mechanical switching, and energy transport. Here, we propose and experimentally demonstrate a nanoscale acoustic lens comprised of 11 gold nanodisks whose collective oscillation gives rise to an interference pattern that results in a diffraction-limited surface acoustic beam of about 340 nm width, with an amplitude contrast of 60%. Via spatially decoupled pump-probe experiments, we were able to map the radiated acoustic energy in the proximity of the focal area, obtaining a very good agreement with the continuum elastic theory.

The transduction of electrical signals into acoustic waves and vice versa is currently used for high-fidelity signal processing in compact systems such as handheld mobile devices. However, modern electronic circuits pose a speed barrier in the few GHz, limiting the possibility of addressing the increasing society's demand for faster technologies. Nanoscale photonic systems are being investigated for new applications in the range of tens to hundreds of gigahertz and beyond to overcome the existing limits.[1–3] In equivalence to its electronic analogue, integrated photonic circuits would also benefit from bidirectional light-hypersonic wave conversion for high-frequency signal processing.[4]

In this scenario, plasmonic nanoantennas emerge as promising candidates for the optical generation and detection of mechanical waves.[5–7] Nanostructured metals act as both optic and mechanical resonators, enabling efficient bidirectional electromagnetic–acoustic wave transduction.[8] When pumped by an intense light pulse at interband transitions, plasmonic nanoantennas produce a large population of hot electrons that ultimately relax by transferring energy to the metal crystal lattice, leading to mechanical vibrations of the nanostructure in its normal modes.[9,10] Such oscillations will contain information embedded in the optical pump pulse,[11,12] which can be emitted as acoustic waves through the environment of the nanoantenna. Placing a second (*receiver*) nanoantenna in the acoustic far field of the *emitter*, the impinging traveling mechanical waves can be converted back to an optical signal. The shape distortions of a vibrating metal nanostructure perturb its plasmonic resonance spectrum, modulating its optical response when probed by a reading light beam.[13,14]

The use of hypersonic waves as mediators between optical signals at the nanoscale has been recently demonstrated experimentally, considering different nanoantenna geometries, laying the basis for photonic-phononic signal processing.[15] Furthermore, by adequately tuning the shape of the resonators, directional control of the acoustic wave emission can, in principle, be achieved. Indeed, using nanorod and V-shape antennas, two-way emission along a single spatial axis has been shown.[14,13] However, up until now, single-way directional control or any other form of acoustic wave confinement using nanosized resonators has not yet been experimentally proven, although a few alternatives have been theoretically proposed,[13,16] and designs of larger structures have been experimentally tested.[17] In this work, we demonstrate the first plasmonic nanoscale hypersonic wave lens, capable of transducing visible light into surface acoustic waves (SAWs) on the order of 10 GHz and focusing them at a controllable distance in the far field. The configuration is evaluated in a dual-color pump-probe experiment, revealing an acoustic focal spot width of about 340 nm, as probed with receiver nanoantennas placed at different locations around the lens. The experimental results are supported by frequency-domain numerical simulations based on the finite element method (FEM), showing very good agreement.

**Results and Discussion**

The design of the devised hypersonic lens is shown in Figure 1a, together with a representative scanning electron microscope (SEM) image in Figure 1b (see fabrication details in Methods section). The device is composed by 11 85-nm diameter, 30-nm height gold disks arranged in the shape of a circular arc of approximately 1.4 µm radius on a quartz substrate. The acoustic lens is designed to operate following Huygens' principle, such that the SAWs emitted by the individual particles interfere throughout the substrate to confine the acoustic field at the focal area. Nanorod gold detectors are placed in the vicinity of the focal point to assess the performance of the lens, with a separate lens-detector pair in the sample for each receiver location. Figure 1, in particular, shows the configuration with the detector placed at the geometric focal point.

The fabricated systems were studied through ultrafast pump-probe measurements using a 400 nm wavelength pump pulse to excite gold interband transitions, and an 800 nm probe pulse near the plasmonic resonance to read the mechanical vibrations of the nanostructures (see further measurement specifics in Methods section). Experiments were carried out either by pumping and probing over the same structure (disk array or rod nanoantenna) to characterize the intrinsic response of each component, or by pumping the lens and probing over a rod detector to evaluate the capabilities of the acoustic lens.

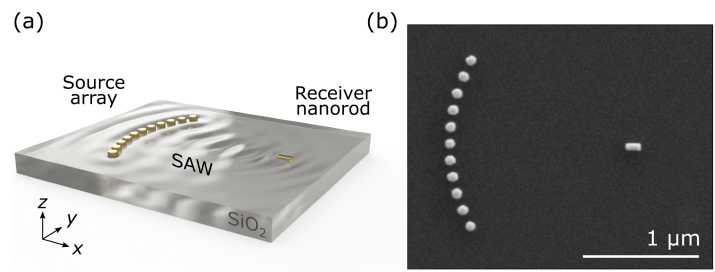

**Figure 1.** Sample design. (a) Sketch of the emitter-array and receiver-nanorod configuration. The emitter antennas are optically excited through the gold interband transition, launching a field of surface acoustic waves (SAWs) through the substrate, then detected by a single gold nanorod located in the far field. (b) SEM image of an emitter-receiver pair where the receiver nanostructure is located at the geometric focal point of the lens, ~1.4 μm apart from its center. The array disks have an average diameter of (85 ± 6) nm, while the receiver rod has an average length of (132 ± 5) nm and a width of (60 ± 3) nm. The dimensions were obtained from SEM images of ~30 emitter-receiver pairs.

Figure 2 (top panels) shows the independently measured pump-probe differential transmission (ΔT/T) temporal traces of the nanodisk array (Figure 2a) and a nanorod antenna (Figure 2b). In both cases, the signal starts with a sudden increase in the magnitude of Δ$T/T$ due to impulsive (electronic) and displacive (lattice expansion) excitation mechanisms, followed by an exponential thermal decay displaying clear oscillations originating from coherent acoustic phonons.[10,18,19] Figure 2c,d presents the simulated absorption cross section of the nanostructures, superimposed with the pump and probe spectra. When the resonators are excited to a dilated shape by the pump pulse, their plasmonic resonances redshift, momentarily increasing (reducing) the nanodisk array (nanorod) probe absorption, as the probe spectrum seats at lower (higher) wavelengths to the resonance absorption peak. This explains the observed negative (positive) sign of the corresponding Δ$T/T$ signal. When the nanoantennas vibrate while relaxing to thermal equilibrium, their plasmonic resonances shift back and forth in wavelength at the same pace, enabling optical detection of the mechanical modes with the probe pulse. Nanorods have been chosen as receiver nanostructures because they present a strong modulation on their optical response when mechanically excited through the underlying substrate.[14,15] In addition, their plasmonic resonance can be accurately tailored by modifying their aspect ratio, enhancing the detection sensitivity.[20] These structures were oriented perpendicular to the axis of the lens to maximize the coupling to surface waves, as previously noted by Imade *et al*.[14] An attempt was made to use nanodisks as receivers, but their low transient optical modulation made it difficult to observe the phonon contribution to the signal.

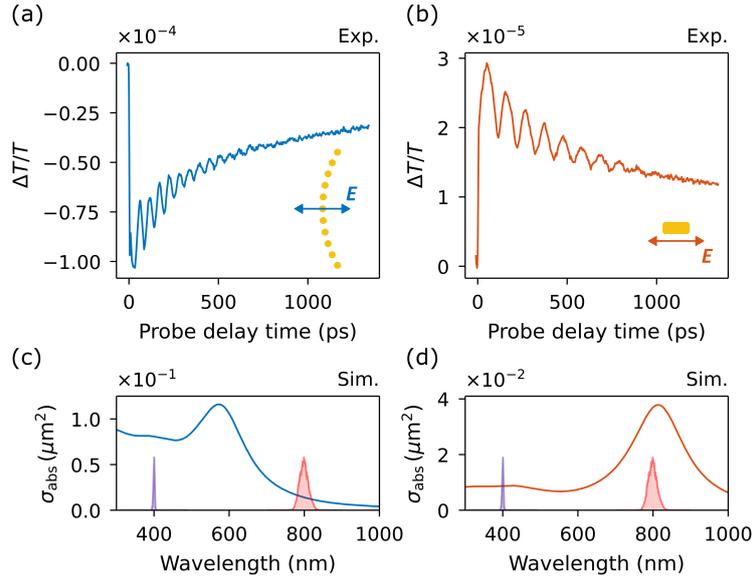

**Figure 2.** Local excitation and detection of coherent acoustic phonons. (a,b) Experimental differential probe transmission signals (Δ*T*/*T*) of the nanodisk array (a) and a single nanorod (b). The arrow in the inset indicates the probe polarization direction. (c,d) Simulated absorption cross section of the emitter and receiver nanostructures for the same incident polarization as indicated in (a,b). The curve filled in red (purple) corresponds to the probe (pump) spectrum.

Figure 3b presents the pump-probe measurement results of the mechanically coupled nanodisk array-nanorod system, with the configuration having the receiver nanoantenna placed at the nominal focal position (sketch in Figure 3a). To reduce variations arising from optical alignment and dispersion in size, shape, and adhesion of the nanostructures to the substrate, we averaged over a statistically significant number (at least 10) of nominally identical lens-receptor pairs for each receptor position (see Supporting Information for more details). The main graph in Figure 3b shows the probe transmission temporal trace and its frequency spectrum when pumping the lens and probing over the nanorod detector. The bottom and lateral panels exhibit the time and frequency projections of the main plot, respectively. As can be observed in the graph at the bottom, the signal sees its temporal onset at 430 ps pump-probe delay time. This is the time it takes for the acoustic wave to travel from the nanodisk array to the nanorod at a speed of 3140 m/s, consistent with the expected value for the Rayleigh wave velocity in a quartz substrate.[21] The weaker response, starting at shorter times, originates from the tail of the probe beam Gaussian spot reaching the acoustic lens, enabling the observation of its vibrational modes. The spectral composition in frequency of the signal (left panel) presents a main peak at 9.5 GHz, corresponding to the extensional mode of the nanorod, and a secondary minor peak at 18.5 GHz, associated to the radial breathing mode of the nanodisks, as discussed next.

To extract the relevant parameters from the experimental temporal traces such as amplitude and frequency, we fitted the curves (as the one displayed in Figure 3b, bottom panel) using the following expression:

$$\frac{\Delta T}{T}(t) = a_0\left[1 + e^{-b(t-t_0)}\right]^{-1} e^{-c_0(t-t_0)} \sin(2\pi f_0(t-t_0) + p_0) + a_1 e^{-c_1 t} \sin(2\pi f_1 t + p_1) \qquad (1)$$

where the first term corresponds to the SAW-driven motion of the receiver (present after the arrival time of the wave), and the second one relates to the optically excited phonons of the lens (present throughout the entire measurement). These two mechanical oscillatory components were accounted for by using exponentially damped sinusoids of amplitude $a_i$, damping constant $c_i$, frequency $f_i$, and phase $p_i$ ($i = 0, 1$). As seen from Eq. (1), the transient impulsive arrival of the acoustic wave is modeled by a sigmoid factor $s(t) = [1 + e^{-b(t-t_0)}]^{-1}$, where $b$ is the steepness coefficient and $t_0$ the inflection point. The arrival time, $t_a$, is taken as the time at which the tangent line to the curve $s(t)$ at the inflection point $t_0$ intercepts the $\Delta T/T = 0$ axis: $t_a = t_0 - 2/b$. Further information regarding the signal processing can be found in the Supporting Information.

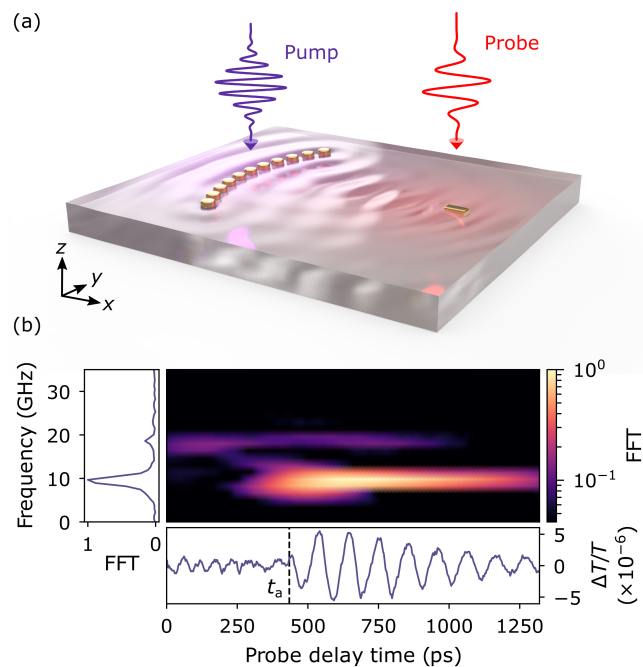

**Figure 3.** Generation and detection of surface acoustic waves. (a) Illustration of the experimental pump-probe configuration employed for the generation and detection of SAWs. The pump beam (400 nm wavelength) illuminates the lens array while the probe beam (800 nm wavelength) is temporally delayed and focused onto the receiver antenna to read its mechanical motion exerted through the substrate. (b) Time-frequency representation of an emitter-receiver pair measurement, where the receiver antenna is located at the geometric focal point, ~1.4 μm apart from the central disk of the array. The vertical dashed line in the bottom panel indicates the arrival time, $t_a$, of the acoustic perturbation. The left panel displays the corresponding fast Fourier transform (FFT) where two peaks can be distinguished. The main peak, at 9.5 GHz, corresponds to the delayed mechanical excitation of the receiver antenna due to the arriving SAWs, whereas the peak at 18.5 GHz arises from the direct optical excitation of the acoustic lens coherent phonons due to a non-negligible component of the probe spot illuminating the structure.

To gain further insight into the measured response of the combined lens-receptor system, we analyze the mechanical frequency spectrum of the constituent subsystems (lens and receptor), as well as that of the quartz substrate, which acts as the coupling medium transporting the hypersonic waves. Figure 4a,d shows the experimental and simulation results for the nanodisk array, respectively (see Methods for details on numerical calculations). In the simulation, the average displacement of the particles along the polarization direction (x-axis) is shown. In both graphs, a strong signal is observed at around 18 GHz, at the radial breathing mode of the nanodisks (labeled as number 2 in the plot), with weaker contributions at lower and higher frequencies (modes labeled as 1, 3, and 4). The spatial deformation distributions of the different modes are shown in Figure 4g. No significant mechanical coupling of acoustic modes between neighboring nanodisks was observed, as expected for a cluster with degenerate modes.[22] Figure 4b,e exhibits the measured and computed response of the bare nanorod, displaying an intense peak around 9 GHz assigned to the extensional mode (labeled as 5), and a smaller contribution at 18 GHz arising from a breathing-like mode (labeled as number 6) along the short axis ($y$-axis direction), as deduced from the numerical results.

Having characterized the intrinsic response of the nanostructures, in Figure 4c we evaluate the behavior of the lens (pump) - detector (probe) configuration with the rod placed at the focal spot (same data as Figure 3b, left panel). The strongest signal occurs at the main natural frequency of the rod detector (9.5 GHz, mode number 5), accompanied by a small unintentional contribution at 18.5 GHz (mode number 2) coming from weakly probing the nanodisk array, as previously discussed. The inset of the graph displays the spectral content of the SAWs arriving from the lens through the substrate, revealing their ability to mechanically excite the nanorod main eigenmode (mode number 5). The substrate displacement, shown in the inset, presents a relative minimum at the maximum frequency component of the lens (mode number 2), where the disks absorb most of the energy, with relative maxima at both higher and lower frequencies (labeled as 8 and 7, respectively), one of them close to the rod extensional mode (mode number 5), indicating a near-optimal scheme. Thus, the mechanical spectral response of the receiver nanoantenna, which depends for instance on its shape and orientation, constrains the detection bandwidth. The Rayleigh-SAW nature of the propagating hypersonic waves can be observed in the simulations of Figure 4f, where the substrate deformation is concentrated only at the surface.

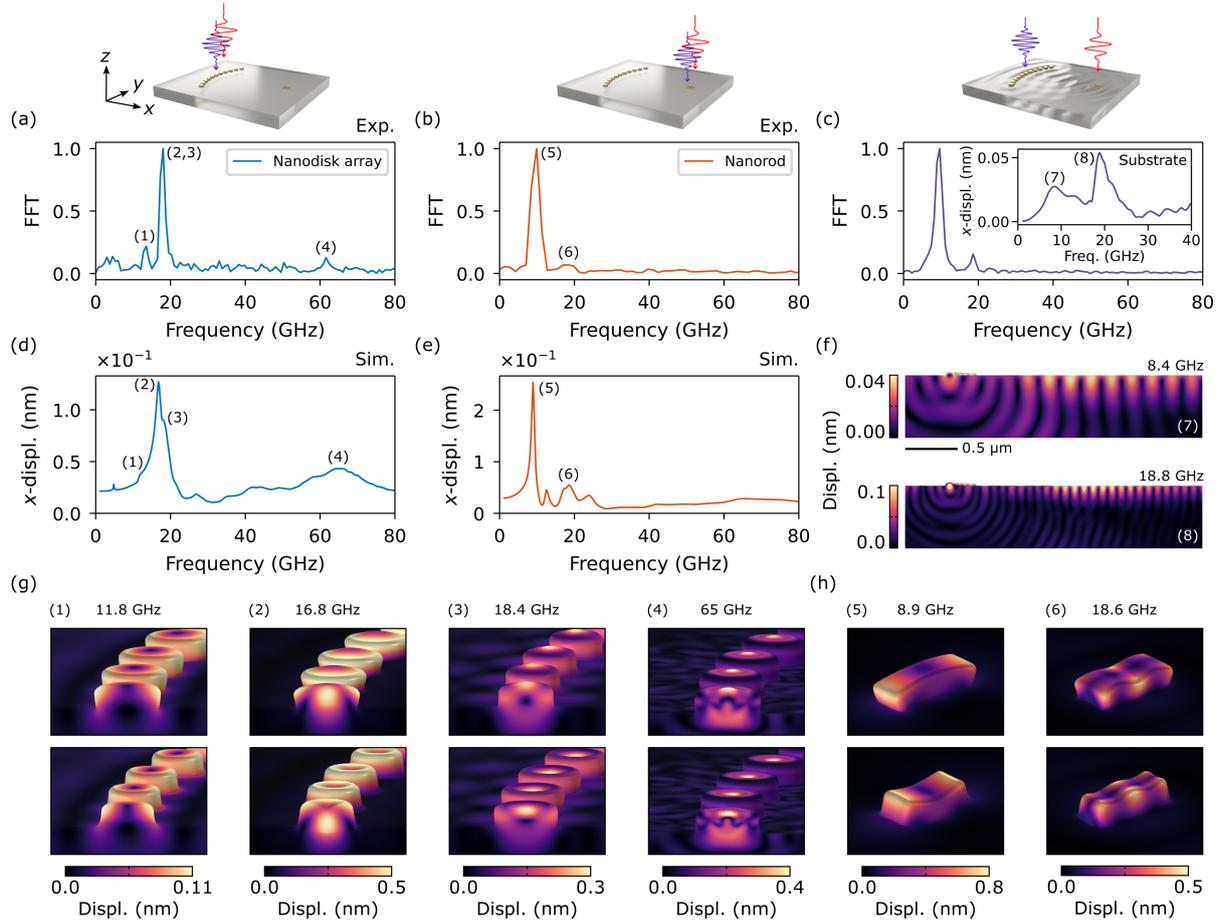

**Figure 4.** Mechanical spectral response. (a,b) FFT of the experimental pump-probe signals shown in Figure 2a,b. (c) FFT of the experimental temporal trace shown in Figure 3b (bottom panel). The inset shows the numerical simulation of the substrate surface displacement at the rod-receptor position, in the direction of the symmetry axis (*x*-axis). The sketches at the top of panels of (a-c) illustrate the pump (purple) and probe (red) spot relative positions. (d,e) FEM frequency-domain simulation of the displacement of the disk array (d) and the nanorod (e) along the symmetry axis. (f) Cross section of the substrate total displacement at the symmetry plane for the frequencies indicated in the inset of panel (c). (g,h) Calculated deformation geometries at the main mechanical resonances of the acoustic lens (g) and the rod detector (h), indicated in panels (d) and (e), respectively. A scale factor of 15x for (g) and 20x for (h) was applied to highlight the deformation.

Finally, we map the acoustic field in the vicinity of the focal region by characterizing multiple lens-detector pairs, with receivers placed at 15 different positions, as schematized in Figure 5a. In the sample, each lens-detector set is spatially separated from each other, to avoid undesired optomechanical coupling. Figure 5b,c shows simulation results of different views of the mechanical displacement of the substrate exerted by the acoustic lens vibrating at the detector main frequency (9.5 GHz), with the arrows marking the different sample sections evaluated in the experiment. Figure 5d-e presents the measured amplitude profile (dots) of the differential probe transmission signal for lens-receptor configurations with detectors placed at different distances from the lens, showing very good agreement with the numerical simulation of the root mean square (RMS) displacement in the steady state (full line). The full width at half maximum at the focal plane is about 340 nm, which slightly increases at the evaluated planes further away. This value is comparable to the Abbe diffraction limit, calculated as $d = \frac{\lambda_{\text{SAW}}}{2\sin\alpha} = \frac{c_{\text{SAW}}}{2f_{\text{SAW}}\sin\alpha} \approx 340$ nm, where $\lambda_{\text{SAW}}$ is the wavelength of the detected acoustic waves, $c_{\text{SAW}} = 3.14$ nm/ps its velocity (see Supporting Information), $f_{\text{SAW}} = 9.5$ GHz its frequency, and $\alpha = 29$ deg is half of the opening angle of the lens. On the other hand, the device achieved an axial amplitude contrast of $(a_{\max} - a_{\min})/(a_{\max} + a_{\min}) \sim 0.6$ within the 660-nm length range measured over the focal plane $X_1$ (Figure 5a), with $a_{\max}$ and $a_{\min}$ representing the maximum and minimum measured amplitudes, respectively.

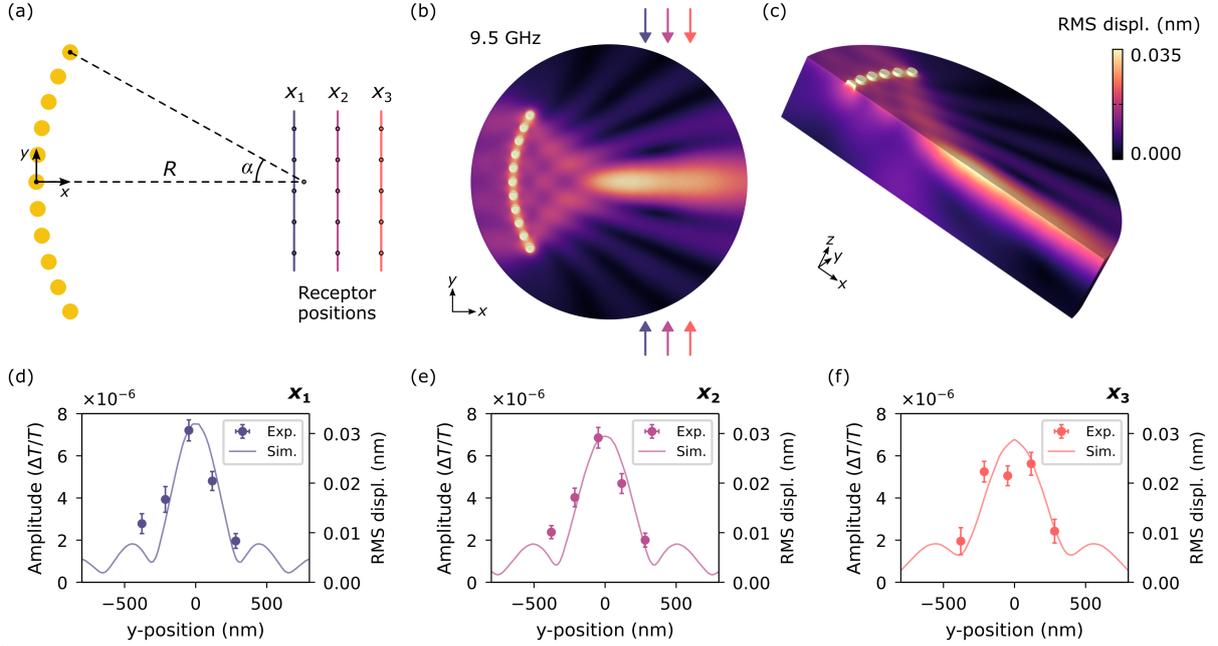

**Figure 5.** Mapping of the acoustic field. (a) Scale diagram of the acoustic lens and the relative rod-receiver positions around the geometric focal point. The lens characteristic dimensions, the radius $R = (1.41 \pm 0.08)$ µm, and the opening angle $\alpha = (29 \pm 1)$ deg, as well as the receiver positions coordinates $X_1 = (1.36 \pm 0.01)$ µm, $X_2 = (1.59 \pm 0.01)$ µm, $X_3 = (1.82 \pm 0.01)$ µm, were estimated from SEM images of the sample. (b,c) Frequency-domain simulation of the substrate in-plane (b) and out-of-plane (c) RMS displacement at the resonant frequency of the rod-detector (9.5 GHz). Arrows in panel (b) indicate the axes where the receivers are positioned. The radius of the simulated domain was taken equal to the lens radius, $R$. (d-f) Mean amplitude of the differential probe transmission signals (circles) compared to the calculated substrate RMS displacement amplitude (solid lines), for the axes indicated in (a).

For the previous analysis, we assume that the modulation amplitude of the optical response is proportional to the deformation amplitude of the receiver structure, and this to the SAW amplitude. The first hypothesis is supported by previous reports showing a linear relationship between the plasmon shift and the deformation amplitude,[5,23] while the latter holds for linear elastic systems. It is worth to mention that the orientation of the nanorod with respect to the SAW propagation direction for the different positions displayed in Figure 5a may introduce slight deviations when comparing the measured amplitudes to the calculated ones.

**Conclusions**

To summarize, we have designed and experimentally demonstrated a hypersonic lens nanostructure capable of focusing surface acoustic waves into a diffraction-limited region of a few hundred nanometers. The mechanical performance of the system has been modeled using FEM linear elastic simulations, obtaining a very good agreement with the experimental results. In this regard, this work opens the possibility of engineering plasmonic nanostructures for the controlled emission of hypersonic acoustic waves at the nanoscale. Furthermore, we show that plasmonic nanoantennas can act as efficient optomechanical transducers to detect minute mechanical vibrations with high optical sensitivity and spatial resolution. By performing spatially decoupled pump-probe experiments, we were able to evaluate the acoustic wave propagation by placing a receiver antenna at different relative positions to the lens to read the substrate deformation in the far field, at the detector resonant frequency.

We anticipate the possibility of obtaining higher-frequency and tighter-focused acoustic fields by reducing the size of the structures that comprise the acoustic lens. Furthermore, enhancement of the focusing intensity can be achieved by optimizing the acoustic lens structure through more complex designs. These devices can be easily integrated into photonic circuits and may be of interest for applications such as directional SAW-driven electron and phonon transport,[24] sensing of mechanical properties in the hypersonic regime,[25,26] mechanical actuation,[27] and nanometric surface acoustic microscopy.[28]

**Methods**

Sample fabrication

The plasmonic nanostructures were fabricated by electron-beam lithography on a quartz substrate. First, the substrate was coated with PMMA resist on which the shape of the antennas was defined. Then, the sample was covered with a 2-nm thick Cr adhesion layer and a 30-nm thick Au film by electron-beam evaporation. The Cr layer reinforces the mechanical coupling of the plasmonic nanostructures to the supporting substrate.[29] Finally, acetone was used to lift-off the resist and the excess of metal.

Pump-probe experiments

Transient transmission measurements were carried out using an ultrafast two-color pump-probe setup with a mode-locked Ti:Sapphire laser (KMLabs). The laser produces ~100 fs pulses with a repetition rate of 95 MHz and an average output power of 300 mW, centered at 800 nm wavelength. The output beam is focused on a BBO crystal, and its second harmonic at 400 nm is modulated at 100 kHz with an acousto-optic modulator and then used as the pump beam. The residual fundamental light, centered at 800 nm, is delayed with a translation stage and used as the probe beam. The measurements were performed using lock-in detection, which allows accurate measurements with a noise floor of $\Delta T/T \sim 10^{-6}$. Both beams were focused onto the sample through the same objective (NA = 0.6) with a spot radius ($e^{-2}$) of 1.1 µm (pump) and 0.8 µm (probe) and were independently directed to different nanostructures by adjusting the angle of incidence, with the help of a home-made dark-field microscope. The average pump and probe powers at the sample were 250 µW and 10 µW, respectively.

Numerical simulations

Numerical calculations were performed using the Structural Mechanics module of the FEM solver software COMSOL Multiphysics. The linear elastic response of the system was obtained by solving the Navier's equation in the frequency domain. A thermal strain $\varepsilon_{\text{th}}$, proportional to the increase in the lattice temperature, $\Delta \tau_l \sim 100$ K, was considered as the displacive excitation mechanism: $\varepsilon_{\text{th}} = \alpha \Delta \tau_l$, where $\alpha$ is the coefficient of linear thermal expansion. To account for the intrinsic damping mechanisms, an isotropic loss factor $\eta = 0.1$ was implemented in the gold domains. Perfectly matched layers (PMLs) were used to truncate the computational domain, simulating an infinite substrate. These artificial domains are not displayed in Figure 4 and Figure 5 for clarity. Further information on the modeling and numerical calculations can be found in Ref. [8].

**Supporting Information**
Size and shape distributions of the fabricated nanostructures, signal processing, experimental signals at all receiver positions, arrival times and SAWs velocity, backward emission of acoustic waves, and sweep in the number of disks comprising the lens.

**Acknowledgments**
This work was partially supported by PICT-2021-I-A-00363, PICT-2021-GRF-TI-00349, PIP 112 202001 01465, UBACyT No 20020220200078BA. A.V.B. acknowledges Alexander von Humboldt Foundation support. L.N. and E.C. acknowledge funding from the European Commission through the ERC Starting Grant CATALIGHT (802989) and DAAD (57573042). The authors acknowledge funding and support from the Deutsche Forschungsgemeinschaft (DFG, German Research Foundation) under Germany´s Excellence Strategy — EXC 2089/1–390776260 e-conversion cluster, the Bavarian program Solar Energies Go Hybrid (SolTech) and the Center for NanoScience (CeNS).

**Conflict of Interest**
The authors declare no conflict of interest.

**Supporting Information for:**

**Focusing Surface Acoustic Waves with a Plasmonic Hypersonic Lens**

Hilario D. Boggiano[1], Lin Nan[2], Gustavo Grinblat[1,3], Stefan A. Maier[4,5], Emiliano Cortés[2], and Andrea V. Bragas*[1,3]

[1] Universidad de Buenos Aires, Facultad de Ciencias Exactas y Naturales, Departamento de Física. 1428 Buenos Aires, Argentina.

[2] Chair in Hybrid Nanosystems, Nanoinstitute Munich, Faculty of Physics, Ludwig-Maximilians-Universität München, 80539 München, Germany

[3] CONICET - Universidad de Buenos Aires, Instituto de Física de Buenos Aires (IFIBA). 1428 Buenos Aires, Argentina.

[4] School of Physics and Astronomy, Monash University, Clayton, VIC 3800, Australia

[5] Department of Physics, Imperial College London, London SW7 2AZ, UK

* bragas@df.uba.ar

Content:

- S1. Size and shape distributions of the fabricated nanostructures.
- S2. Signal processing.
- S3. Experimental signals at all receiver positions.
- S4. Arrival times and SAWs velocity.
- S5. Backward emission of acoustic waves.
- S6. Sweep in the number of disks comprising the lens.

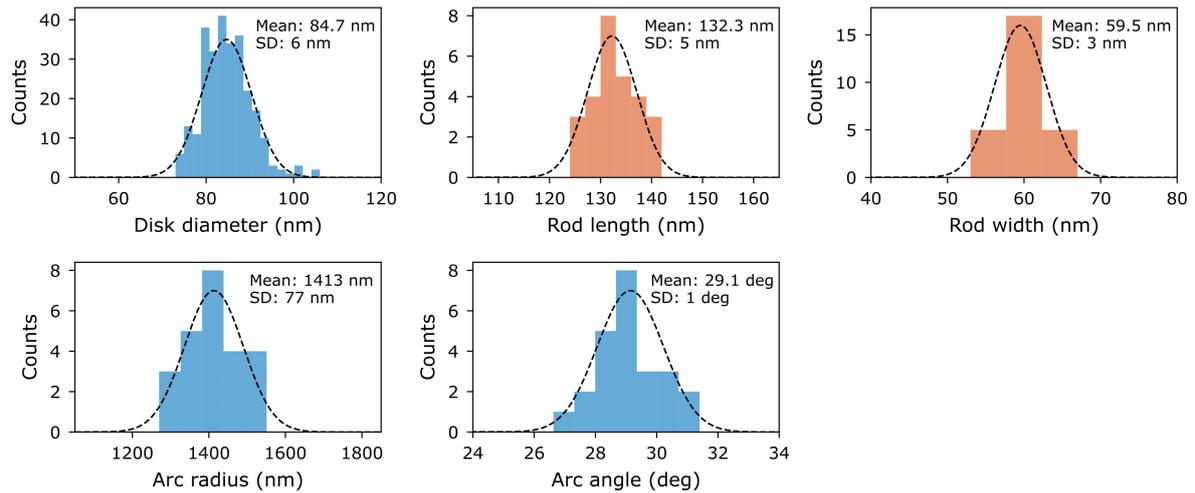

**Figure S1.** Size and shape distributions of the fabricated nanostructures. The size of the gold nanostructures as well as the dimensions of the lens were estimated from SEM images of about 30 emitter-receiver pairs. The height of the gold structures is 30 nm, with a 2-nm thick Cr layer underneath to improve adhesion to the quartz substrate. The distribution's mean value and standard deviation (SD) are displayed in the inset of each panel.

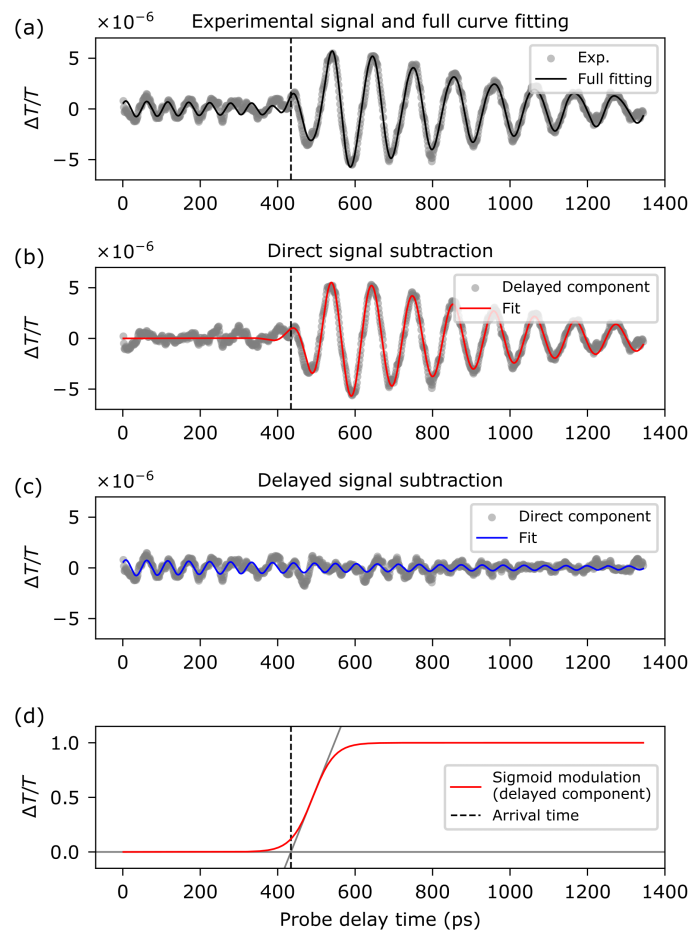

**Figure S2.** Signal processing. (a) Experimental data (same as in Figure 3b, bottom panel) and the corresponding fitting curve following the model presented in the main text (Eq. 1). This model considers a "delayed component" associated with the SAW-driven motion of the receiver nanorod, and a "direct component" associated with the optically excited motion of the acoustic lens. (b,c) Decomposition of the experimental data and fitting curve into the delayed (b) and direct (c) components. (d) Sigmoid modulation factor, $s(t)$, of the delayed signal, which models the impulsive arrival of the acoustic wave. The arrival time of the perturbation (vertical dashed line) is taken as the time at which the tangent line to the curve $s(t)$ at the inflection point intersects the $\Delta T/T = 0$ axis.

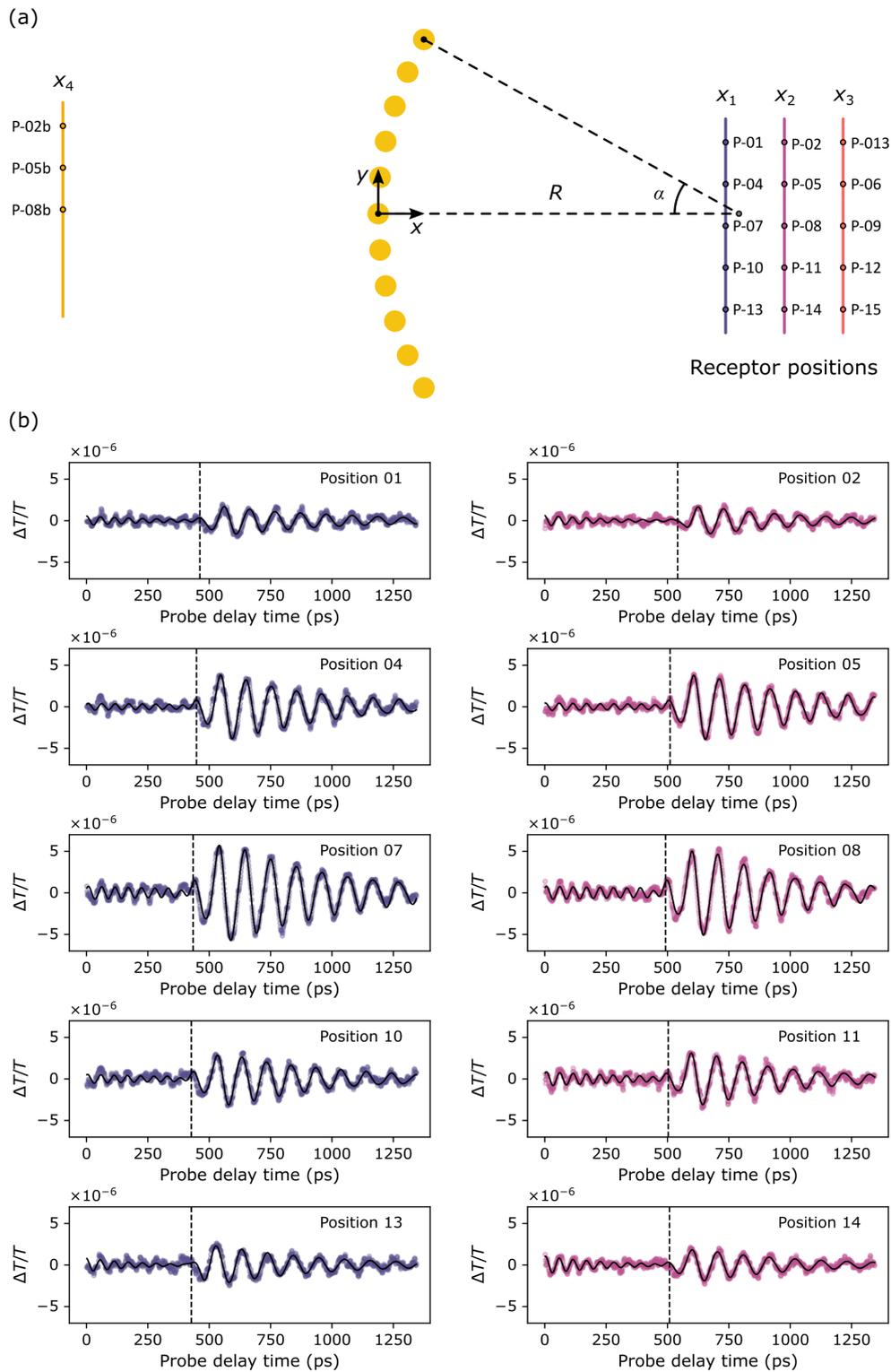

**Figure S3.1.** Experimental signals at all receiver positions. (a) Scale diagram of the acoustic lens and the relative rod-receiver positions. $R = (1.41 \pm 0.08)$ μm, $\alpha = (29 \pm 1)$ deg, $X_1 = (1.36 \pm 0.01)$ μm, $X_2 = (1.59 \pm 0.01)$ μm, $X_3 =$

($1.82 \pm 0.01$) μm, $X_4$ = ($1.24 \pm 0.01$) μm. (b) Experimental differential probe transmission signals ($\Delta T/T$) of different lens-detector pairs, with receivers placed at different positions, as illustrated in (a).

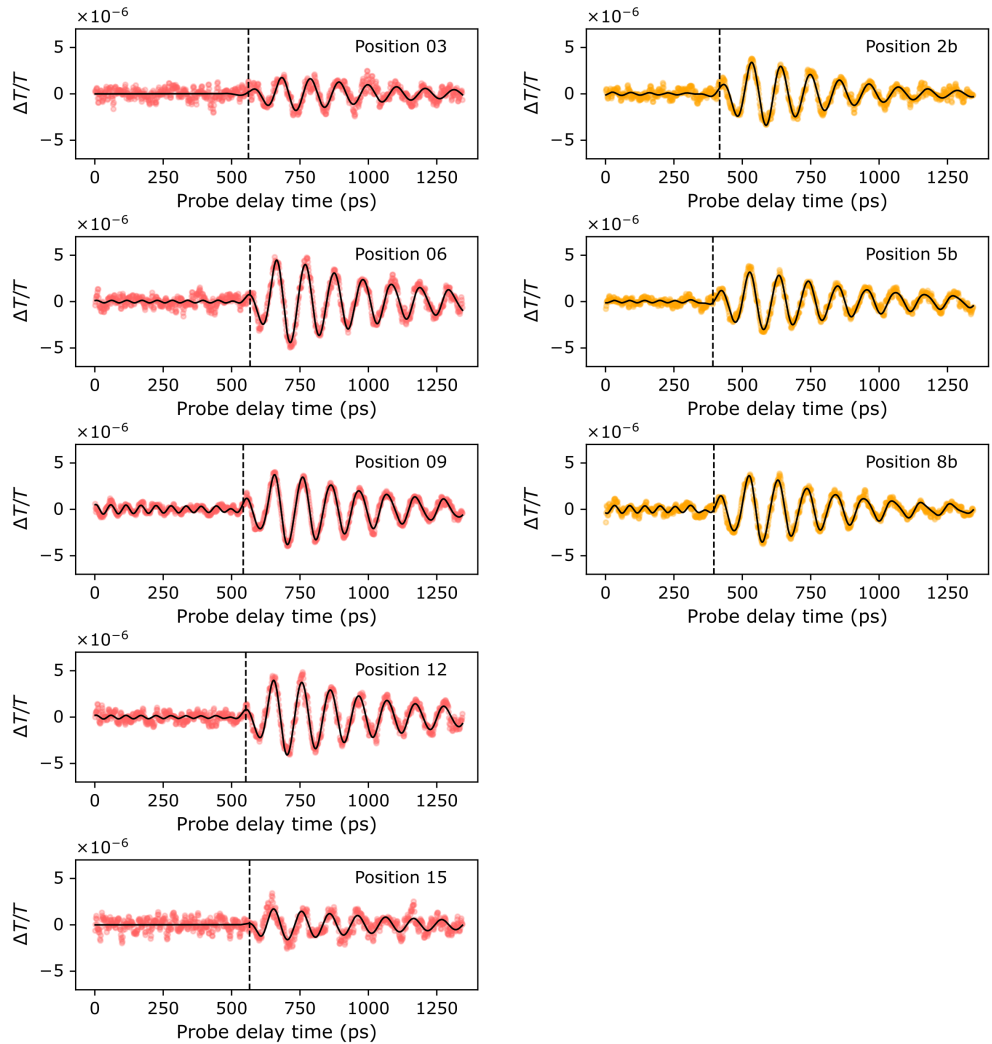

**Figure S3.2.** As Figure S3.1b for the rest of the receiver positions.

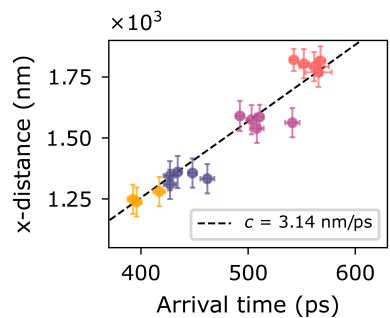

**Figure S4.** Arrival times and SAWs velocity. Minimum lens-receiver distance along the *x*-axis (*x*-distance) direction as function of the measured arrival time. Each experimental data point corresponds to a given receiver position (see Figure S3.1a). The SAW velocity, obtained from the linear fit of the data (dashed line), is 3.14 nm/ps.

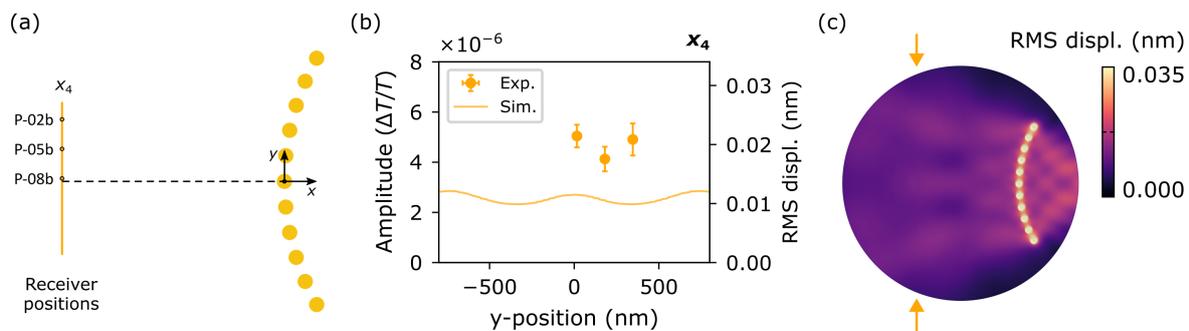

**Figure S5.** Backward emission of acoustic waves. (a) Sketch of the acoustic lens and the relative receiver positions on the back. $X_4$ = -(1.24 ± 0.01) μm. (b) Mean amplitude of the differential probe transmission signals (circles) compared to the calculated substrate RMS displacement amplitude (solid line). (c) Frequency-domain simulation of the substrate RMS-displacement at the resonant frequency of the receiver (9.5 GHz). The arrows mark the sample section evaluated in the experiment.

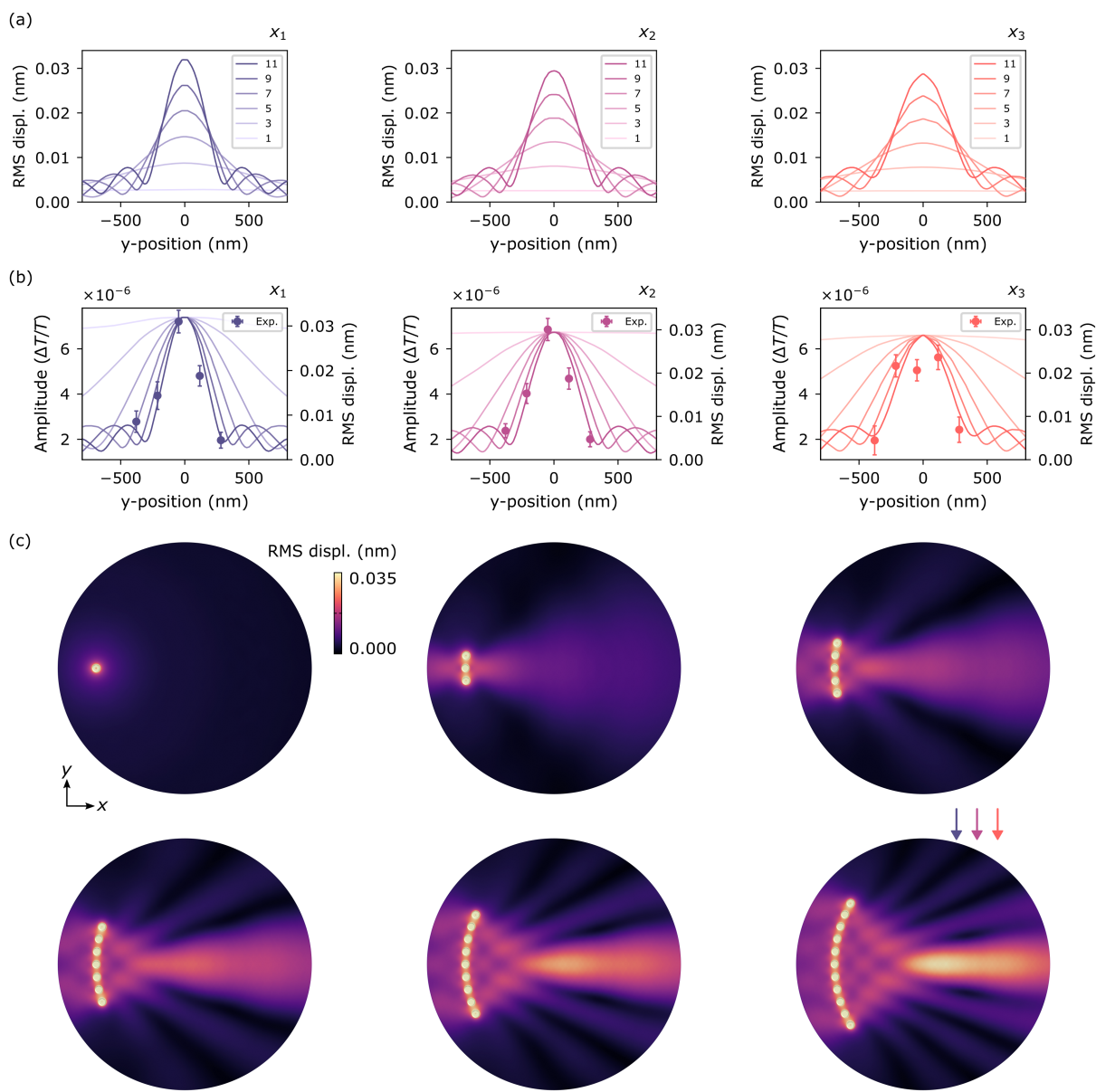

**Figure S6.** Sweep in the number of disks comprising the lens. (a) Simulation of the substrate displacement amplitude at the axes indicated in Figure 5 of the main text for different numbers of gold nanodisks comprising the acoustic lens (indicated in the inset of each panel). (b) Comparison between the relative amplitude of the differential probe transmission signals (circles) and the normalized amplitude distribution obtained in the simulations (same data as in Figure 5), showing very good agreement when considering the full lens structure (11 disks). (c) Substrate displacement patterns. Arrows in the bottom-right panel indicate the axes $X_1$, $X_2$, and $X_3$, where the receivers are located. The radius of the simulation domain is $R = 1.41$ µm.